\documentclass{PoS}

\title{Non-standard heavy vectors at the LHC}

\ShortTitle{Non-standard heavy vectors at the LHC}

\author{\speaker{Gennaro Corcella}\\
  INFN, Laboratori Nazionali di Frascati\\
  Via E.~Fermi 40, 00044 Frascati (RM), Italy\\
        E-mail: \email{gennaro.corcella@lnf.infn.it}}

\abstract{Heavy vector-boson hunting is one of the major analyses undertaken
by the  experiments carried out at the LHC.
I will explore two scenarios which have not been investigated yet by
the experimental collaborations.
The first scenario consists in searching for $Z'$ bosons,
predicted by U(1)$'$
GUT-inspired models, by studying its 
supersymmetric decays,
leading to resonant final states with charged leptons and
missing energy. Afterwards, I shall discuss the 331 model and the
production of
doubly-charged vector bileptons, yielding events with
two same-sign lepton pairs at the LHC.}

\FullConference{The European Physical Society Conference on High Energy Physics\\
		5-12 July, 2017\\
		Venice}

\begin{document}

\section{Introduction}
Hunting for heavy vectors is one of the challenging goals
of the LHC collaborations,
since such vectors may represent a possible evidence of physics beyond the 
Standard Model.
Heavy neutral gauge bosons $Z'$ are predicted in extensions of
the Standard Model (SM) based on U(1)$'$ symmetries inspired
by Grand Unification Theories (GUT) \cite{langa}
and in the Sequential Standard Model (SSM), the simplest extension of the
SM, wherein $Z'$ and $W'$ bosons have the same coupling
to fermions as $Z$ and $W$.
The LHC experiments have so far assumed that the $Z'$ has only SM
decay modes and have searched for high-mass dileptons
\cite{atlasll,cmsll} and dijets \cite{atlasjj,cmsjj},
setting exclusion limits on the
$Z'$ mass in the multi-TeV range.

In this talk, I explore possible supersymmetric decays of 
GUT-inspired $Z'$ bosons, within the Minimal Supersymmetric
Standard Model (MSSM), along the lines of 
Refs.~\cite{corgen,cor}.
The opening of new channels lowers the 
SM branching ratios and the exclusion
limits; moreover, if a $Z'$ were to be discovered, its decays
would be an ideal
place to look for supersymmetry, since the $Z'$ mass
would set a constraint on sparticle invariant masses.
In particular, decays into the lightest neutralinos are
useful for the purpose of 
Dark Matter searches at colliders.

The other scenario which I shall tackle is the
so-called 331 model, in the formulation of \cite{frampton},
which features five additional gauge bosons, 
four charged and one neutral, plus
three exotic quarks. 
One of the main signatures of this model consists of
the production of doubly-charged vector bileptons, decaying into
same-sign charged-lepton pairs.
ATLAS and CMS did explore
doubly-charged Higgs bosons $H^{\pm\pm}$ \cite{hh1,hh2}, i.e. scalar bileptons, 
setting model-dependent exclusion bounds of several hundreds of GeV,
but never searched for vector bileptons.
In this talk, I will investigate processes
with two vector bileptons and two jets at the LHC, and compare their
signal with SM backgrounds.

Section 2 will be devoted to $Z'$ bosons within supersymmetry, discussing
first the theoretical framework and then the phenomenology, while
Section 3 will deal with the 331 model and bilepton production at the
LHC. I will finally make some concluding remarks in Section 4.

\section{Supersymmetry and $Z'$ bosons}
\subsection{Theoretical framework}
U(1)$'$ symmetries and $Z'$ bosons typically arise from the breaking of a 
rank-6  GUT group ${\rm E}_6$ :
\begin{equation}\label{e6}
{\rm E}_6\to {\rm SO}(10)\times {\rm U}(1)'_\psi\to 
{\rm SU}(5)\times {\rm U}(1)'_\chi \times {\rm U}(1)'_\psi.
\end{equation}
A generic $Z'$ is given by the mixing between
$Z'_\psi$ and $Z'_\chi$, associated with
U(1)$'_\psi$ and U(1)$'_\chi$:
\begin{equation}\label{ztheta}
Z'(\theta)=Z'_\psi\cos\theta-Z'_\chi\sin\theta.
\end{equation}
This talk will be concentrated on the $Z'_\psi$ and its supersymmetric
decays,
since
it is the framework yielding the
largest cross section \cite{cor}.
Scenarios with other values of $\theta$ 
were detailed in
\cite{corgen,cor}.

In the MSSM, besides the scalar Higgs doublets $H_d$
and $H_u$, a singlet $S$ is necessary
to break the U(1)$'$ symmetry and give mass to the $Z'$.
After electroweak symmetry breaking, the Higgs sector consists of
one pseudoscalar $A$ and three scalars $h$, $H$ and a new $H'$.
In the gaugino sector, one will have 
two extra neutralinos, whereas the charginos
are unchanged; furthermore, 
the U(1)$'$ group leads to extra D- and F-term
corrections to sfermion masses.

\subsection{Phenomenology of supersymmetric $Z'$ decays}

Following \cite{cor}, I shall discuss the phenomenology of a
$Z'_\psi$ boson with mass 
$m_{Z'}=2~{\rm TeV}$ and require that the coupling
constants of U(1)$'$ and U(1)$_{\rm Y}$ are
proportional via $g'=\sqrt{5/3}~g_1$.
As for the MSSM parameters, I shall set:
$M_1=400$~GeV, $M'=1$~TeV, $\tan\beta$=30,
$\mu=200$~GeV and
$A_q=A_\ell=A_\lambda=4$~TeV, where $A_q$ and $A_\ell$ are the soft
trilinear couplings of squarks and sleptons with the Higgs fields
and $A_\lambda$ is the soft Higgs trilinear coupling.

The sfermion spectrum, corresponding to soft masses at the $Z'$ scale
$m_{\tilde\ell}^0=m_{\tilde\nu_\ell}^0=1.2$~TeV and 
$m^0_{\tilde q}$=5.5~TeV, computed by means of the SARAH \cite{sarah}
and SPheno \cite{spheno} codes,
is given in the tables reported in \cite{cor} and is omitted here
for the sake of brevity.
In the reference point of Ref.~\cite{cor}, 
the branching ratio
into supersymmetric final states is almost
30\%: the highest rate is into chargino pairs 
$\tilde\chi^+_1\tilde\chi^-_1$, accounting
for about 10\%, while the one into the
lightest neutralinos, Dark Matter candidates, is
roughly 5\%.
An interesting supersymmetric channel can therefore be:
\begin{equation}\label{chain}
pp\to Z'_\psi\to\tilde\chi_1^+\tilde\chi_1^-\to
  (\tilde\chi_1^0\ell^+\nu_\ell)(\tilde\chi_1^0\ell^-\bar\nu_\ell),
\end{equation}
  with $\ell=\mu, e$, leading to
two charged leptons and missing energy.
Its cross section, calculated at leading order (LO) 
by MadGraph \cite{madgraph}, is given by $7.9\times 10^{-4}$~pb
at 14 TeV.

\begin{figure}[t]
\centerline{\resizebox{0.4\textwidth}{!}
{\includegraphics{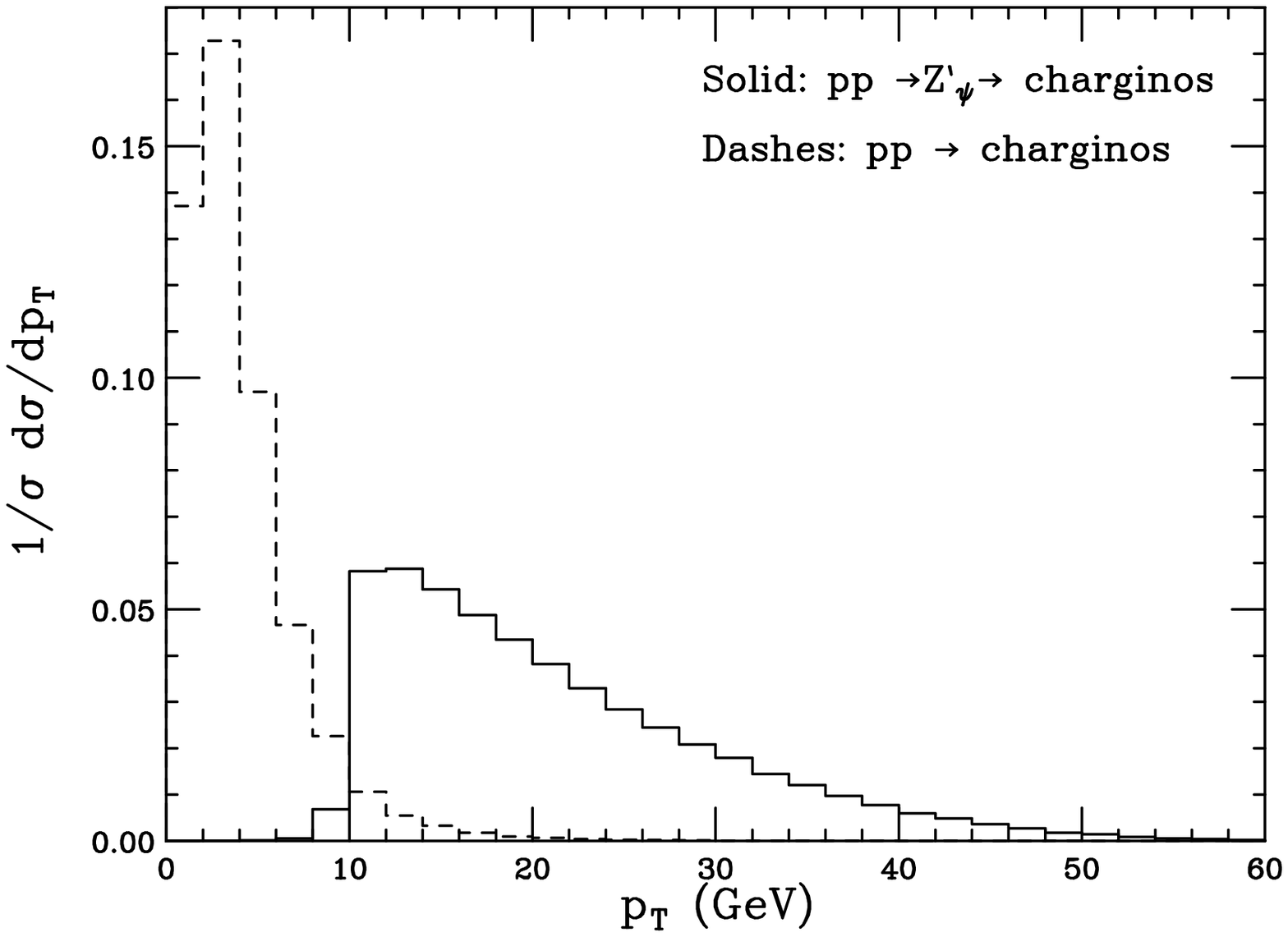}}%
\hfill%
\resizebox{0.4\textwidth}{!}{\includegraphics{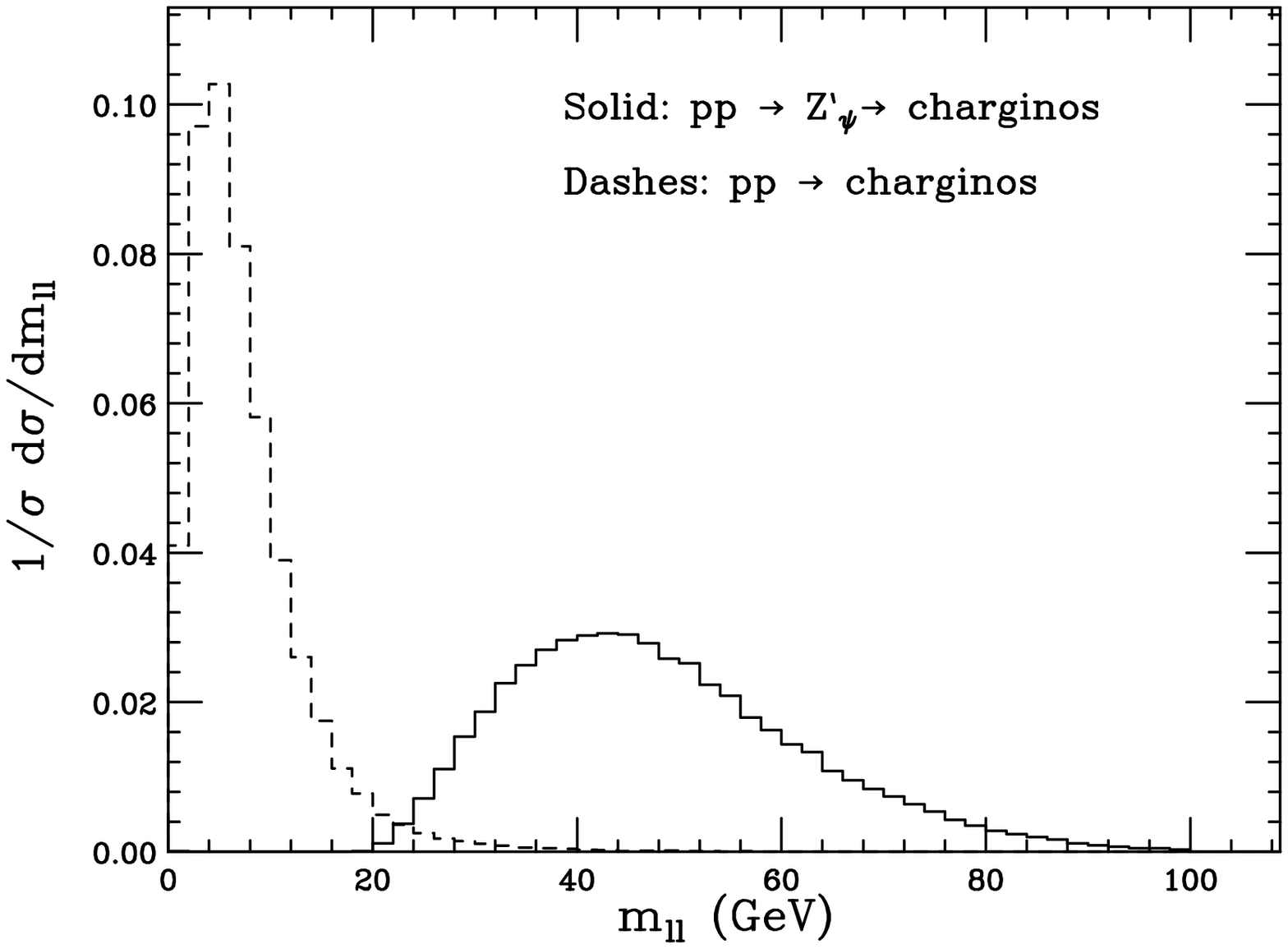}}}
\caption{Left: lepton transverse momentum
  for the $Z'_\psi$ model at $\sqrt{s}=14$~TeV in 
$Z'_\psi\to\tilde\chi^+_1\chi^-_1$ events and direct 
chargino-production processes. Right: invariant mass of the
two charged leptons.}
\label{zpsipt}
\end{figure}

Figure~\ref{zpsipt} (left)
presents the transverse-momentum spectrum of the leptons
produced in Eq.~(\ref{chain}), according to MadGraph+HERWIG \cite{herwig},
as well as in processes with direct
chargino production, i.e. $pp\to\tilde\chi^+_1\tilde\chi^-_1$, followed
by $\tilde\chi^+\to \tilde\chi_1^0\ell^+\nu_\ell$.
For direct chargino production,
the leptons are rather soft and the spectrum is peaked at low $p_T$;
as for $Z'$-driven charginos, 
the $\tilde\chi^+_1\tilde\chi^-_1$
invariant mass must be equal to $m_{Z'}$, and therefore the
lepton-$p_T$ distribution is substantial at higher $p_T$
and can be easily discriminated.
The right panel of Fig.~\ref{zpsipt} displays the $\ell^+\ell^-$ invariant-mass
distribution: for $Z'$ decays into charginos, $m_{\ell\ell}$ 
lies in the range 20 GeV~$<m_{\ell\ell}<$~100 GeV and is
maximum at $m_{\ell\ell}\simeq 45$~GeV;
for direct $\tilde\chi_1^+\tilde\chi_1^-$ production, $m_{\ell\ell}$ 
is peaked about 5 GeV and rapidly vanishes.
Decays into neutralino pairs $Z'\to\tilde\chi^0_1\tilde\chi^0_1$
are relevant for the searches
for Dark Matter candidates: it was shown in \cite{cor} that,
although the shape of the missing-energy spectrum
is similar to the one due to decays into neutrinos, the overall
production  cross section of invisible particles is remarkably
increased, once accounting for neutralinos.
As a whole, even though a specific analysis on supersymmetric
$Z'$ models is currently missing,
using the LHC data at 8 TeV and a benchmark point of the
MSSM parameter space, Ref.~\cite{lhcp} proved that 
the impact of the inclusion of supersymmetric decays
on the mass exclusion limits is about $200-300$~GeV.

\section{Bilepton signature at the LHC}

\subsection{Theoretical framework}

The gauge structure of the bilepton model \cite{frampton},
also known as 331 model,
is $SU(3)_c \times SU(3)_L \times U(1)_X$.
It is a quite appealing framework, as it treats the third quark family
in a different
way with respect to the other two families and predicts
anomaly cancellation
between families, and not for each generation, as long as the
numbers of colours and families are equal.
As for quarks, three exotic states are predicted:
$D$ and $S$, having charge $-4/3$ and lepton number
$L=+2$, and $T$, with charge
$5/3$ and $L=-2$.
Moreover, the model accommodates four neutral Higgs bosons,
three scalar ($h_1$, $h_2$ and $h_3$)
and one pseudoscalar ($h_1$), along with two singly-charged
($h_1^{\pm}$, $h_2^{\pm}$) and one doubly-charged ($h_1^{\pm\pm}$)
Higgses. In the vector sector, a $Z'$ is also predicted, as well as
singly- and doubly-charged bileptons $Y^\pm$ and
$Y^{\pm\pm}$, with $L=2$ ($Y^{++}$ and $Y^+$) or $L=-2$
($Y^{--}$ and $Y^-$).

\subsection{Phenomenology of bileptons at the LHC}
Production of $Y^{++}Y^{--}$ pairs in Drell--Yan processes
were investigated in \cite{nepo} and an exclusion limit about
$m_{Y^{\pm\pm}}>850$~GeV was set, after comparing
the bilepton prediction with the ATLAS search for
doubly-charged Higgs bosons \cite{hh1}.
In this talk, I will instead review the main results in
\cite{cccf}, where the authors explored 
bilepton pairs accompanied by two jets:
\begin{equation}\label{yyjj}
  pp\to Y^{++}Y^{--}jj\to (\ell^+\ell^+)(\ell^-\ell^-)jj.
  \end{equation}
The analysis in \cite{cccf} features a benchmark point where the
bileptons $Y^{\pm\pm}$ have mass about 873 GeV, the $Z'$ is leptophobic
and with $m_{Z'}\simeq$~3 TeV, the exotic quarks are at
$1.6-1.7$~TeV and the new Higgs bosons are too heavy to contribute to
bilepton phenomenology.
In the reference point of \cite{cccf}, the LO cross section
of process (\ref{yyjj}),
calculated by MadGraph,
reads, at $\sqrt{s}=13$~TeV:
$\sigma(pp\to Y^{++}Y^{--}jj\to 4\ell jj)
\simeq 3.7$~fb. Moreover, the following acceptance cuts are imposed:
$p_{T,j}>30$~GeV, $p_{T,\ell}>20$~GeV, $|\eta_j|<4.5$,
$|\eta_\ell|<2.5$, $\Delta R_{jj}>0.4$,  $\Delta R_{\ell\ell}>0.1$,
$\Delta R_{j\ell}>0.4$.

\begin{figure}[t]
\centerline{\resizebox{0.4\textwidth}{!}
{\includegraphics{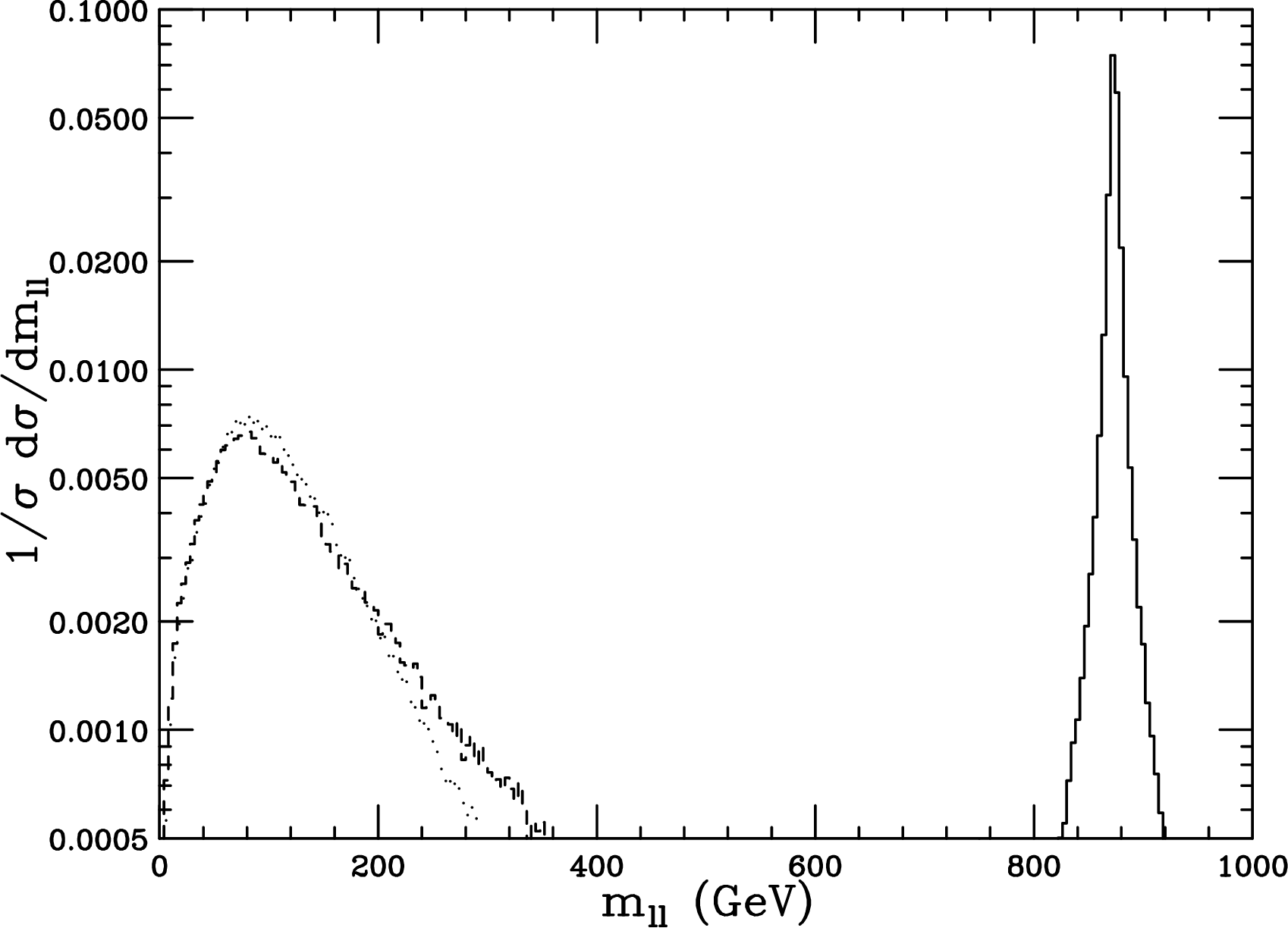}}%
\hfill%
\resizebox{0.4\textwidth}{!}{\includegraphics{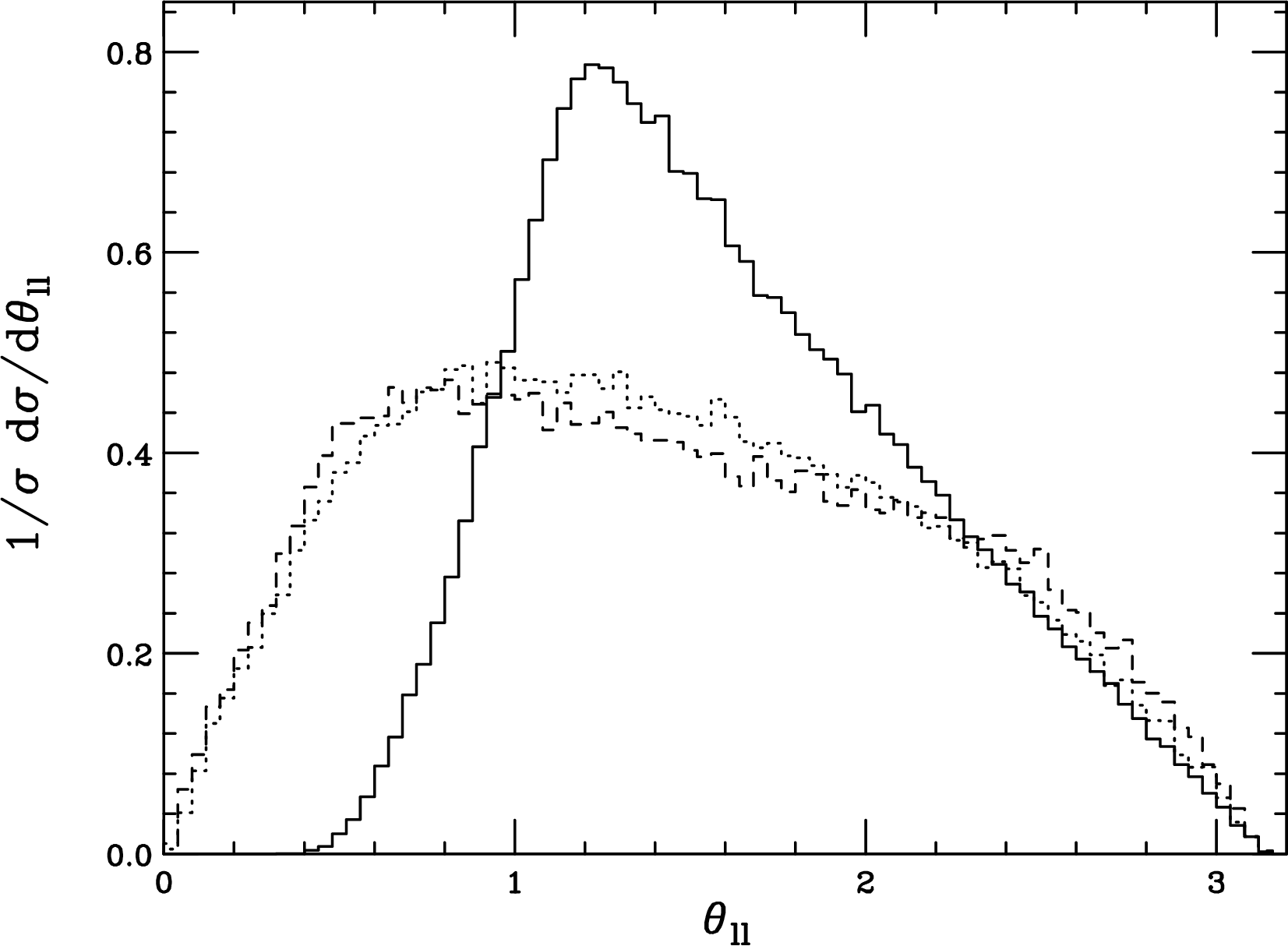}}}
\caption{Left: invariant-mass distribution of same-sign lepton
  pairs. Right: opening angle between same-sign leptons.
  Plotted are the spectra yielded by our signal (solid),
  $ZZjj$ (dashes) and $t\bar tZ$ (dots) backgrounds.}
\label{fig331}
\end{figure}

In Fig.~\ref{fig331} (left) I present the invariant-mass spectrum
  of same-sign lepton pairs (left) in our signal (solid histogram)
  and in the background processes $ZZjj$ (dashes) and $t\bar tZ$ (dots),
  according to MadGraph+HERWIG;
  in the right panel of Fig.~\ref{fig331}, $\theta_{\ell\ell}$ is instead
  the angle between same-sign leptons
  in signal and background processes.
From Fig.~\ref{fig331} one learns that 
our bilepton signal can be
easily separated from the backgrounds.
The invariant mass $m_{\ell\ell}$
distributions are indeed very different: the signal peaks at
$m_{Y^{++}}\simeq 873$~GeV and manifests as a narrow resonance,
whereas the backgrounds
yield broader spectra, peaked around 80 GeV and vanishing
for $m_{\ell\ell}> 350$~GeV.
Regarding the angular distributions, our signal predicts a higher event fraction
at $1<\theta_{\ell\ell}<2$ with respect 
to all backgrounds.

\section{Conclusions}
I investigated two scenarios featuring heavy vectors at the LHC,
which have not yet received adequate consideration from the experimental
collaborations, namely supersymmetric decays of GUT-inspired $Z'$
bosons and doubly-charged vector bilepton production.
I demonstrated that primary $Z'$ decays into
charginos lead to final states with charged leptons and missing energy, which
can be discriminated from direct non-resonant chargino production. 
Likewise, the production of vector-bilepton pairs and
two jets can yield a visible signal at the LHC, which can be separated from the
backgrounds setting typical acceptance cuts.
It is now certainly worthwhile reconsidering the
investigation in \cite{nepo} on vector $Y^{++}Y^{--}$
production in Drell--Yan processes,
so as to understand how the final-state distributions fare with respect
to scalar $H^{++}H^{--}$ production, studied in \cite{hh1,hh2}, and
used in \cite{nepo} to set the exclusion limits
on the $Y^{\pm\pm}$ mass. This is in progress.

\end{document}